# Efficient Ambient LoRa Backscatter with On-Off Keying Modulation

Xiuzhen Guo, *Student Member, IEEE,* Longfei Shangguan, *Member, IEEE,* Yuan He*, *Senior Member, IEEE,* Jia Zhang, *Student Member, IEEE,* Haotian Jiang, *Student Member, IEEE,* Awais Ahmad Siddiqi, *Student Member, IEEE,* and Yunhao Liu, *Fellow, IEEE/ACM*

*Abstract*—Backscatter communication holds potential for ubiquitous and low-cost connectivity among low-power IoT devices. To avoid interference between the carrier signal and the backscatter signal, recent works propose a frequency-shifting technique to separate these two signals in the frequency domain. Such proposals, however, have to occupy the precious wireless spectrum that is already overcrowded, and increase the power, cost, and complexity of the backscatter tag. In this paper, we revisit the classic ON-OFF Keying (OOK) modulation and propose Aloba, a backscatter system that takes the ambient LoRa transmissions as the excitation and piggybacks the in-band OOK modulated signals over the LoRa transmissions. Our design enables the backsactter signal to work in the same frequency band of the carrier signal, meanwhile achieving flexible data rate at different transmission range. The key contributions of Aloba include: i) the design of a low-power backscatter tag that can pick up the ambient LoRa signals from other signals; ii) a novel decoding algorithm to demodulate both the carrier signal and the backscatter signal from their superposition. We further adopt link coding mechanism and interleave operation to enhance the reliability of backscatter signal decoding. We implement Aloba and conduct head-to-head comparison with the state-of-the-art LoRa backscatter system PLoRa in various settings. The experiment results show Aloba can achieve 39.5–199.4 **Kbps data rate at various distances,** 10.4–52.4× **higher than PLoRa.**

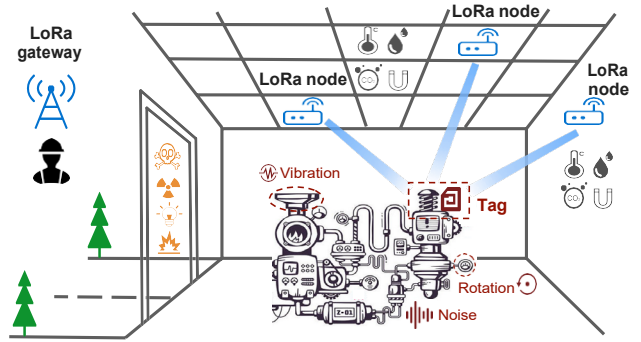

Fig. 1. An illustration of Aloba deployment in the factory. Aloba tag takes the ambient LoRa signals as the excitation and transmits the machine status data (*e.g.*, vibration) back to the gateway hundreds of meters away.

## I. INTRODUCTION

The fourth industry revolution (*a.k.a.* industrial 4.0) aims to transform traditional manufactory and industrial practices with advanced automation, artificial intelligence, and the internet of things technology. The key to the success of industrial 4.0 is the underlying machine to machine (M2M) communication technology that provides ubiquitous and reliable wireless connectivity anywhere, at any time [1], [2], [3], [4], [5], [6]. However, industrial applications have diverse requirements on data exchanging and thus demand different M2M communication technologies. For example, video surveillance on industrial practices requires high-throughput (tens of Gbps) M2M links to ensure reduced communication latency [7]. Keeping track of the status of machines in the factory (*e.g.*, vibration, noise, rotation), on the other hand, demands moderate-throughput (hundreds of Kbps) M2M links for sensing data forwarding [8]. As some of these machines may produce strong noise, flash intense light, and discharge harmful gases (as shown in Figure 1), these data forwarding links should be also low-power and long-range, allowing sensors to transmit their data back to the gateway hundreds of meters away without extra human intervention or frequent battery replacement.

While the latest 5G new radio [9] and 802.11ax (*a.k.a.* Wi-Fi 6) [10] technologies have been successfully deployed in factories for low-latency data transmission, these technologies are not suitable for machine monitoring in the complex industrial environment as they are either susceptible to blockage or constrained by short communication range (tens of meters). On the other hand, Low-power wide-area networks (LPWANs) are able to connect machines across long range. However, these technologies consume substantial amount of energy in around-the-clock monitoring mode and thus require frequent battery replacement.

Due to its low-power and simplicity, backscatter communication becomes a promising technology to enrich the family of existing M2M links. The state-of-the-art backscatter systems (Figure 2) now can transmit at high throughput [11], [12], [13], [14], [15], [16], [17], [18], [19], [20], [21] and communicate over long distance [22], [23], [24]. These desirable properties make backsactter communication a good candidate for industrial applications. However, as we carefully examine these innovations, we find these designs (Figure 2)

Xiuzhen Guo, Yuan He, Jiang Zhang, Haotian Jiang, Awais Ahmad Siddiqi and Yunhao Liu are with the School of Software and BNRist, Tsinghua University, P.R. China. Longfei Shangguan is with University of Pittsburgh and Microsoft.

E-mail: guoxiuzhen94@gmail.com, longfei.shangguan@microsoft.com, heyuan@mail.tsinghua.edu.cn, {j-zhang19, jht19}@mails.tsinghua.edu.cn, xidq19@mails.tsinghua.edu.cn, yunhaoliu@gmail.com

*Yuan He is the corresponding author.

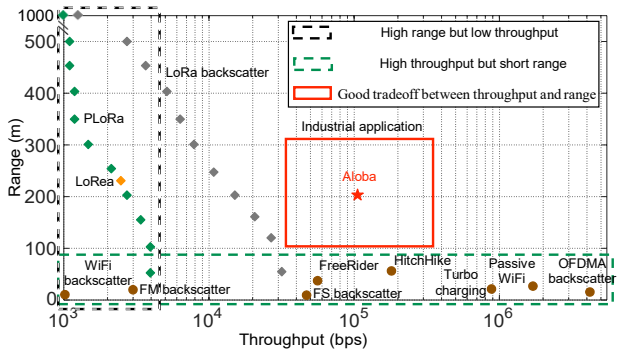

Fig. 2. Comparison of existing backscatter technologies.

either sacrifice the communication range in order to achieve a higher link throughput, or tradeoff the throughput for a decent communication range. None of them is able to balance the throughput and communication range to satisfy the requirement of the machine status monitoring in industrial practices. For example, the throughput of OFDMA-backscatter [21] can reach up to 5.2 Mbps, it however only supports short-range communication (up to 10 m). In contrast, PLoRa [24] supports kilometer-scale backscatter communication at the cost of a very low throughput (limited to tens of Kbps). Similarly, LoRa backscatter [23] allows the backscatter tag to communicate with the gateway kilometers away. However, its maximum link throughput is constrained to 27.3 Kbps due to the LoRa PHY-layer regulation. Besides, LoRa backscatter relies on the dedicated excitation source to send a continuous sinusoidal tone as the carrier signal. This operation inevitably adds cost and complicates the installation and maintenance of the system.

LoRa transmits data using chirp-modulated signals that can be decoded at very low signal-to-noise ratio (SNR), thus in principle serving as an excellent excitation signal for long-range backscatter. This observation leads us to propose Aloba, an ambient backscatter design for machine status monitoring in industrial practices. Aloba takes the ambient LoRa signals (*e.g.*, emitted from nearby active LoRa nodes) as the carrier signal, modulating its sensing data on these carrier signals using ON-OFF Keying (OOK) modulation, and then reflects the modulated signal to the receiver (LoRa gateway). Such design owns three desirable properties: *i*): *Flexible link throughput*. OOK modulation enables the Aloba tag to adapt its link throughput to the link quality as opposed to the PHY-layer regulation of carrier signals. *ii*): *Long communication range*. By taking ambient LoRa signals as the carrier, the Aloba tag could leverage the unique processing gain brought by the chirp signal design to enable backscatter communication at a range of hundreds of meters. *iii*): *Easy to deploy*. The Aloba tag modulates its data on the ambient LoRa signals. This can avoid the installation and maintenance of the dedicated excitation sources.

Harvesting these benefits, however, faces fundamental challenges. On one hand, unlike the conventional RFID system where the excitation is a continuous wave (a sinusoidal tone), the excitation signal in our design is an ambient, intermittent LoRa signal which already conveys information and changes over time. The backscatter tag should be able to distinguish the LoRa signal from the others and further synchronize with each LoRa symbol for fine-grained modulation. On the other hand, the backscatter signal is orders of magnitude weaker than the excitation signal. The LoRa receiver will receive the direct excitation signal from the LoRa transmitter and the backscatter signal from the Aloba tag. In other words, the received signal is the **superposition** of excitation signal and backscatter signal. The receiver should be able to demodulate this weak backscatter signal in the presence of strong interfering excitation signal. Due to the frequency variation, the superposition of an excitation signal and a backscatter signal changes over time, which makes the demodulation even more challenging (§II).

In Aloba we present a novel hardware-software solution to tackle the above challenges. On a high-level the Aloba tag picks up the ambient LoRa signal from the other signals using a low-power LoRa packet detection circuit. It then modulates data on the LoRa payload chirps using OOK. The LoRa receiver leverages our signal processing algorithm to decode both the excitation and the backscatter signals from their superposition. Link coding mechanism and interleave operation are used to enhance the reliability of backscatter signal decoding. Moreover, we discuss the impact of synchronization of carrier signal and backscatter signal on the performance of Aloba and propose moving window-based decoding strategy to tolerant synchronization errors. Aloba makes three key contributions:

- We design a simple yet effective LoRa packet detection circuit that can detect the ambient LoRa signal as low as -60 dBm with 0.3 $mW$ power consumption. This packet detection circuit serves as a plug-in peripheral that can be easily integrated with commercial backscatter tags, *e.g.*, WISP.
- We comprehensively study the superposition of chirp signals at the LoRa receiver, based on which we propose a novel demodulation algorithm that can detect the fine-grained changes on the phase and amplitude of the received signal to demodulate both the carrier signal and the backscatter signal.
- We implement Aloba on a PCB (printed circuit board) and integrate it with WISP [25] for evaluation. The experimental results demonstrate that Aloba can achieve various data rates (39.5–199.4 Kbps) at various tag-to-receiver distances (50–200 m) in the wild, given the tag-to-source distance of 10 m. Compared with the state-of-the-art system PLoRa [24], Aloba achieves 10.4–52.4× higher throughput. We shared our schematic and source code in http://tns.thss.tsinghua.edu.cn/sun/researches/Backscatter Communication.html for reproducibility.

Compared with the published SenSys version [26], we improve the hardware design of Aloba tag and propose a new low-power packet detection circuit in Section IV. We also discuss the power consumption and management of the modified Aloba tag. We add experiments to analyze the distribution of error bits and explain the error types more clearly in Section V. We further propose the link coding mechanism, interleave operation, and coherent combining at the receivers for decoding reliability enhancement. The evaluation results demonstrate the effectiveness of these methods in Section IX. The demodulation extension to the multi-tag scenario is introduced in Section VI. In Section VII, we discuss the impact of synchronization process on the performance of Aloba and propose the potential solutions. Finally, we discuss the challenges in the real application scenarios of Aloba in Section XI. Specifically, we discuss the impact of LoRa duty cycle and industrial environment (especially the vibration, rotation, and EMI generated by the industrial machines) on Aloba decoding in Section XI-A and Section XI-D. We discuss the feedback channel of Aloba and propose the potential solutions to achieve the tradeoff between data rate and communication range in Section XI-B. We also discuss the SNR requirement of Aloba decoding and propose the potential solutions to extend the communication range in Section XI-C.

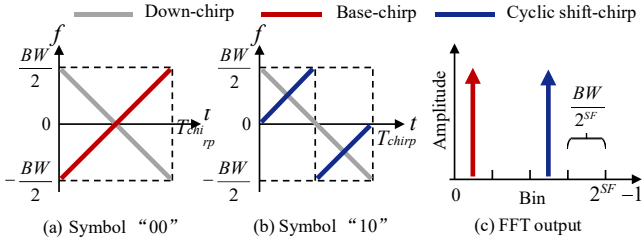

Fig. 3. LoRa modulation and demodulation. (a): the base-chirp encodes symbol "00". (b): a cyclic shifted version of the base-chirp encodes symbol "10". (c): the multiplication of different chirps and down-chirp yields peaks on different FFT bins.

## II. SELF-INTERFERENCE ON LORA BACKSCATTER

Aloba modulates ambient LoRa signal (the carrier signal) using OOK. In this section, we first introduce the standard LoRa modulation and demodulation process, and then analyze LoRa self-interference.

### A. LoRa Primer

LoRa adopts Chirp Spread Spectrum (CSS) to modulate data. Each LoRa symbol is represented by a chirp where the frequency changes linearly over time, as shown in Figure 3(a)-(b). To demodulate the LoRa symbol, the receiver multiplies an incoming LoRa symbol with a down-chirp and transforms the multiplication from the time domain to the frequency domain, yielding a peak on an FFT bin. The receiver tracks the location of this peak to demodulate the LoRa symbol accordingly. Figure 3(c) illustrates this process.

### B. Modeling the Interference

The frequency of a LoRa chirp can be represented by: $f(t) = F_0 + kt$, where $F_0$ is the initial frequency of this LoRa chirp; $k$ is the frequency changing rate (over time). The phase of this LoRa chirp at a given time $t$ can be calculated by:

$$\phi(t) = 2\pi \int_0^t f(t)\, dt = 2\pi(F_0 t + \frac{1}{2}kt^2),\ t \leq T_{chirp} \quad (1)$$

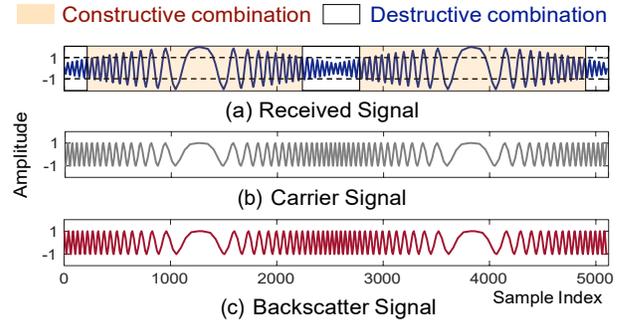

Fig. 4. The received signal is the superposition of the carrier signal and backscatter signal (propagation delay of the backscatter signal is 1.8 μs). The received signal experiences periodic amplitude variation due to constructive and destructive interference.

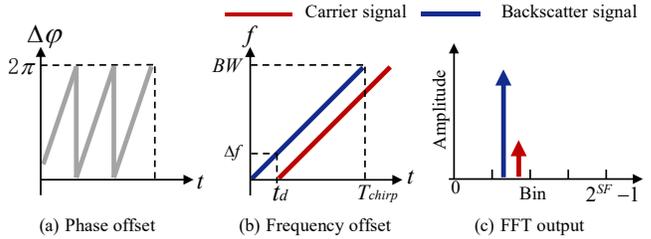

Fig. 5. Frequency-time domain characteristics of chirp signal combination. (a): the phase offset between the carrier and backscatter signal changes over time. (b): the propagation daley $t_d$ between these two signals leads to an offset $\Delta f$ on the frequency domain. (c): both carrier signal and backscatter signal drop to the same FFT bin after demodulation due to the small frequency shift.

**Phase offset.** The backscatter signal and the carrier signal propagate along different paths. The propagation delay due to this path difference $d$ can be represented by $t_d = \frac{d}{c}$, where $c$ is the radio propagation speed. At time $t$, the phase of the carrier signal and the backscatter signal are $\phi_c(t) = 2\pi(F_0 t + \frac{1}{2}kt^2)$ and $\phi_b(t) = 2\pi(F_0(t-t_d) + \frac{1}{2}k(t-t_d)^2)$, respectively. Hence the phase difference $\Delta\phi(t)$ between these two signals can be calculated by:

$$\Delta\phi(t) = \phi_c(t) - \phi_b(t) = 2\pi((F_0 + kt)t_d - \frac{1}{2}kt_d^2) \quad (2)$$

From the above equation we have the following observations: i): The phase difference $\Delta\phi(t)$ varies over time, as shown in Figure 5(a). Hence the received signal experiences interference that periodically alternates between constructive and destructive states, as shown in Figure 4. ii): The propagation delay changes across tag's locations, so does the amplitude variation of the received signal. Therefore, we cannot rely on the amplitude variation of the received signal to detect the appearance of the backscatter signal. On the other hand, conventional interference cancellation algorithms, e.g., passive RFID and Wi-Fi backscatter [27], [28], [29], [15] are not suitable for solving this LoRa self-interference, since the carrier signal here is unknown to the LoRa receiver. Reconstructing the carrier signal in hopes of canceling it out from the received signal is difficult, since the amplitude and frequency of LoRa signals change over time.

**Frequency offset.** The propagation delay $t_d$ leads to a frequency offset in the frequency domain, as shown in Fig-

TABLE I
FREQUENCY OFFSET BETWEEN THE CARRIER SIGNAL AND THE BACKSCATTER SIGNAL UNDER DIFFERENT PARAMETERS

| Path Difference | BW (KHz) | SF | Frequency Offset | FFT Bin |
|---|---|---|---|---|
| 10 m | 500 | 7 | 65.104 Hz | 3906.25 Hz |
| 50 m | 500 | 7 | 325.52 Hz | 3906.25 Hz |
| 100 m | 500 | 7 | 651.04 Hz | 3906.25 Hz |
| 200 m | 500 | 7 | 1302.08 Hz | 3906.25 Hz |
| 400 m | 500 | 7 | 2604.16 Hz | 3906.25 Hz |
| 600 m | 500 | 7 | 3906.25 Hz | 3906.25 Hz |

ure 5(b). The frequency offset $\Delta f$ can be calculated by:

$$\Delta f = \frac{BW}{T_{chirp}} t_d = \frac{BW}{T_{chirp}} \frac{d}{c} = \frac{BW^2}{2^{SF}} \frac{d}{c} \quad (3)$$

The frequency offset is determined by the LoRa bandwidth $BW$, the spreading factor $SF$, and the path difference $d$.

Whereas, the limited resolution of FFT bin at the LoRa receiver makes it hard to apply the existing LoRa parallel decoding methods for Aloba. Table I lists the maximum frequency offset under different path difference settings. We observe that the receiver is unable to differentiate the backscatter signal and carrier signal in frequency domain, unless the path difference is larger than 600 m. As the tag's location is usually unknown in many outdoor deployments, we cannot blindly borrow the idea of LoRa parallel decoding [30], [31], [32] to separate the two signals from their superposition in the frequency domain.

## III. DEMODULATION

This section describes the way to decode both the carrier signal and the backscatter signal from their superposition. Due to signal attenuation, insertion loss, and energy transformation loss on the backscatter tag, the backscatter signal is orders of magnitude weaker than the carrier signal. Hence we can first leverage capture effect to demodulate the carrier signal, using the standard LoRa demodulation scheme. Capture effect means that the stronger of two signals at the same channel can be demodulated from the superposition due to the capture effect [33]. Our task is then transformed into decoding the backcsatter signal from the received signal.

**Basic idea**. The demodulation scheme of Aloba is motivated by the conventional RFID system. RFID system simplifies the backscatter signal decoding by adopting a sinusoidal tone as the carrier signal. In a similar way, if the LoRa receiver can transform the standard LoRa carrier signal into a constant sinusoidal tone, the variation of received signal would be solely determined by the backscatter signal. This implies an opportunity to detect the presence of backscatter signal and then decode it.

### A. Transforming Chirps into a Constant Sinusoidal Tone

**Signal Transformation**. The standard LoRa demodulation scheme multiplies LoRa chirps with a down-chirp (with the frequency changes from $+\frac{BW}{2}$ to $-\frac{BW}{2}$), yielding a sinusoidal tone. The frequency of this sinusoidal tone is determined by the initial frequency offset of LoRa chirps. In Aloba, we replace the standard down-chirp with the *conjugate* of incoming LoRa chirps, as shown in Figure 6(a)-(b). The LoRa chirp and its conjugate chirp are symmetric to each other with respect to the reflection off the X-axis. This operation transforms all incoming LoRa chirps into a sinusoidal tone at the same frequency, as shown in Figure 6(c).

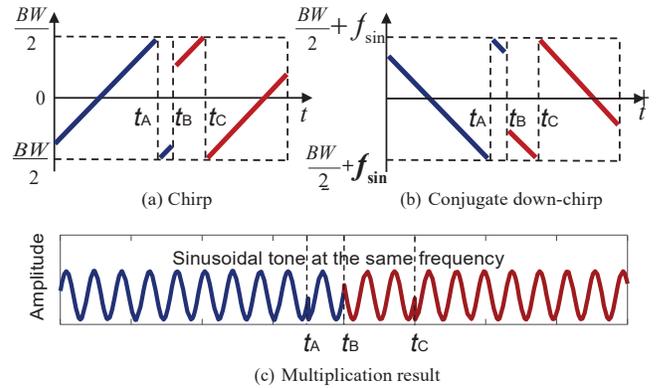

Fig. 6. Signal transformation. (a): two LoRa chirps with different initial frequencies. (b): conjugate down-chirps of these LoRa chirps. (c): the multiplication of the chirps and their conjugate down-chirps yields sinusoidal tone at the same frequency.

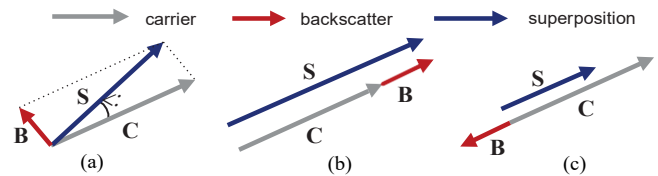

Fig. 7. Superposition of sinusoidal carrier signal and backscatter signal. (a): carrier signal and backscatter signal are not strictly aligned or misaligned. (b): carrier signal and backscatter signal are strictly aligned. (c): carrier signal and backscatter signal are strictly misaligned.

**Amplitude variation of the sinusoidal tone**. The sinusoidal tone obtained from the signal transformation can be regarded as the superposition of a sinusoidal carrier signal and a backscatter signal. As the backscatter signal is much weaker than the carrier signal, the receiver faces two possible signal combinations. *i*): When these two signals are either strictly aligned or misaligned (Figure 7(b)-(c)), we expect to see a significant amplitude variation on the received signal, based on which we can detect the presence of the backscatter signal. *ii*): Most of the time, however, these two signals are neither strictly aligned nor misaligned ((Figure 7(a)). Hence the amplitude of the received signal would not exhibit significant variation in the presence of backscatter signal. As a result, we cannot solely rely on the amplitude variation to detect the backscatter signal in the later case.

**Phase variation of the sinusoidal tone**. We instead leverage the phase variation of the received signal to detect the presence of the backscatter signal in the later case. When Aloba tag is at the OFF state, the received signal is determined by the carrier signal. When Aloba tag switches to the ON state, the presence of backscatter signal will alter the received signal, which leads to a phase jumping on the sinusoidal tone, as shown in Figure 8(a). This phase jumping caused by ON-OFF switching could serve as a clue to detect the presence of the backscatter signal. However, false alarm remains as both the frequency wrapping ( from $\frac{BW}{2}$ to $-\frac{BW}{2}$) within the LoRa chirp and the sinusoidal tone transformation of each LoRa chirp could lead to an abrupt phase jump (denoted as false alarms in Figure 8(a)).

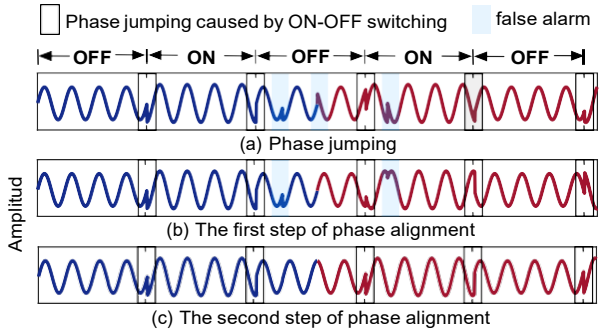

Fig. 8. Phase variation of the received signal after signal transformation. (a): phase jumping caused by ON-OFF switching of backscatter signal and false alarms. (b): eliminate the false alarm caused by signal transformation at the boundary of each LoRa chirp. (c) eliminate the false alarm caused by frequency wrapping within the LoRa chirp.

**Two-step phase alignment**. We design a two-step phase alignment algorithm to eliminate false alarms. In the first step, the receiver checks the boundary of each LoRa symbol on the received signal, wrapping the phase of remaining signal samples for a proper amount of degree such that they are all aligned with the phase samples of the LoRa symbol ahead. This process is repeated from the first LoRa symbol to the last one. After this step, all phase jumpings caused by sinusoidal transformation will be removed, as shown in Figure 8(b). In the second step, the receiver locates those phase jumpings caused by frequency wrapping by reconstructing the carrier symbols.[1] We can also leverage the phase jitter (false phase jumping) proposed in LiteNap [34] as the physical fingerprints to track the continuity of phases before/after frequency wrapping. The receiver further visits these phase jumping points sequentially and repeats the phase alignment operation in step one. After this process, all phase jumpings caused by frequency wrapping will be eliminated from the transformed sinusoidal tone, leaving us true positives (those caused by the presence of backscatter signal) only, as shown in Figure 8(c).

### B. Signal Reconstruction and Decoding

In practice, phase noise exists due to timing offset and carrier frequency offset [35]. It is thus unreliable to tell the presence of backscatter signal solely based on abrupt phase jumping points. To solve this problem, we design a robust, clustering-based detection algorithm as follows.

**Signal reconstruction**. The backscatter signal is modulated on the payload part of the carrier signal (will be detailed in §IV-B). Once the receiver detects the payload of a LoRa packet, the receiver reconstructs the transformed sinusoidal tone as $\mathbf{S} = A_i \Phi_i$, where $A_i$ is the $i^{th}$ amplitude sample on the transformed sinusoidal tone. $\Phi_i$ is the phase difference between the $i^{th}$ phase sample on this transformed sinusoidal tone and the corresponding phase sample on a reference sinusoidal tone, as shown in Figure 10(a). Figure 9 shows the amplitude and phase of the reconstructed signal. We can see the amplitude variation of this reconstructed signal becomes indistinguishable when these two signals are not strictly aligned, while the phase readings of this reconstructed signal demonstrate a distinctive pattern, which can be leveraged for backscatter signal decoding.

---

[1] Since the content of carrier signal has already been decoded, the receiver can directly obtain the location of the frequency wrapping point within each LoRa symbol.

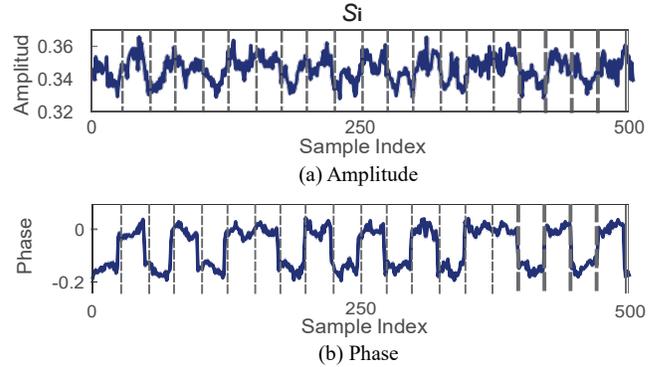

Fig. 9. Reconstructed signal. (a): amplitude variation of the reconstructed signal becomes indistinguishable when carrier signal and backscatter signal are not strictly aligned. (b): the phase of this reconstructed signal shows a distinctive pattern.

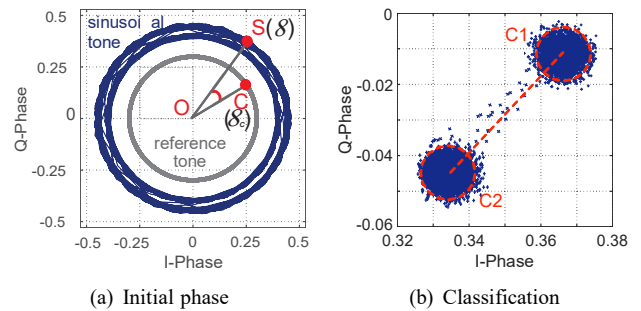

Fig. 10. The illustration of backscatter data decoding. (a): phase difference between the transformed sinusoidal tone plotted by blue lines and a reference sinusoidal tone plotted by gray lines. (b): reconstructed signal samples with and without backscatter signals aggregate two clusters, and backscatter preamble provides classification anchor points.

**Backscatter signal decoding**. We plot all reconstructed signal samples on the constellation diagram (I-Q plane). As shown in Figure 10(b), these symbols are naturally grouped into two clusters: one for the none-existence of backscatter signal, and another for the existence of backscatter signal. We can then leverage the preamble of the backscatter signal (a 16-bit Barker code, detailed in §IV-B) to distinguish these two clusters and decode each trail of backscatter signals accordingly. This clustering-based method leverages all signal samples to detect the presence of backscatter signal, hence it is more robust than the phase jumping based detection method.

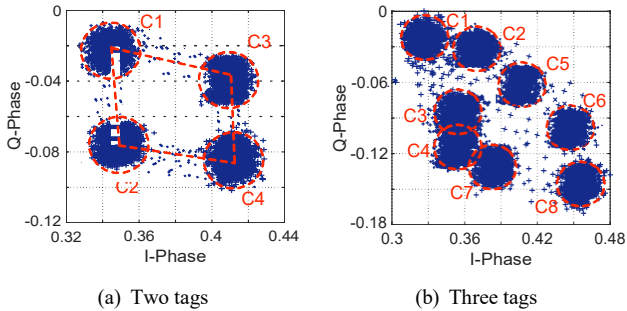

(a) Two tags  (b) Three tags

Fig. 11. Classification results on I-Q plane when multiple tags are concurrent transmitted. (a): two tags correspond to four clusters. (b): three tags correspond to eight clusters.

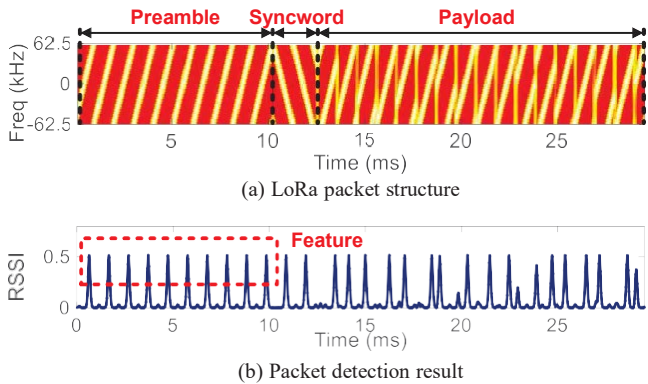

(a) LoRa packet structure

(b) Packet detection result

Fig. 12. LoRa preamble and the corresponding RSSI profile. The ten consecutive up-chirps on LoRa preamble (top) leads to ten equally-spaced RSS pulses (below).

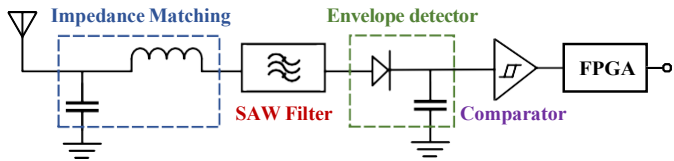

Fig. 13. The circuit design of LoRa packet detector.

On a high-level, Aloba shares the similar decoding principle with the standard LoRa decoding algorithm: transforming the frequency-shifting LoRa chirp into a constant sinusoidal tone. However, the conventional LoRa decoding algorithm multiplies the incoming LoRa chirp with a standard down-chirp and then tracks the peak on FFT bins to demodulate the LoRa chirps. In contrast, the Aloba receiver replaces this down-chirp with the *conjugate* of each incoming LoRa chirp. It then tracks the amplitude and phase variation to demodulate the backscatter signals overlaid on the carrier LoRa signals. This allows the Aloba receiver to decode both the carrier signal and backscatter signal on the same frequency band.

## IV. ALOBA TAG DESIGN

### A. Low-power LoRa Packet Detection

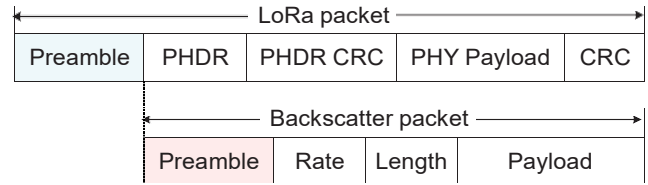

Fig. 14. LoRa packet structure and backscatter packet structure.

signal. The standard LoRa packet detection scheme is not suitable for Aloba due to its high power consumption.

In Aloba, we design a simple yet effective packet detection circuit to pick up the LoRa signal from the ambient noise and unconcerned signals. Our design exploits the unique pattern of LoRa symbols in the LoRa preamble: ten consecutive up-chirps with zero initial frequency offset. When the incoming signal passes through a low-pass filter (with a cutoff frequency at $BW/4$), the ten consecutive up-chirps on a LoRa preamble will lead to ten equally spaced RSS (received signal strength) pulses (Figure 12(b)), whereas the noise and other legacy signals will not. Hence Aloba can pick up the LoRa preamble by detecting the appearance of this unique RSS pattern.

Figure 13 shows the circuit design of Aloba packet detection module. First, we adopt a Surface Acoustic Wave (SAW) filter of Qualcomm B3715 [36] to support narrowband filtering and offers reduced size, weight, and cost, compared to the RF filtering technology. More importantly, the SAW filter is a purely passive component with zero power consumption. Second, the filtered signal is down-converted to the baseband through an envelope detector for demodulation. To minimize the power consumption on detection, an intuitive solution is using a low-power voltage comparator to replace ADC. The comparator (NCS2202) [37] quantizes the RSS signal to High ("1") and Low ("0") two logical voltages. We empirically set this RSS threshold to -60 dBm, which yields the best detection accuracy in our experiments. We then leverage the built-in low-power counter in the FPGA to sample these logical voltages. Once the FPGA detect ten equally spaced RSS pulses, it immediately knows the arrival of a LoRa packet and automatically switches to the modulation mode.

**Power consumption and management**. Both impedance matching, SAW filter and envelope detector are passive components (*e.g.*, inductance, capacitors, and diodes), hence the energy consumption of this packet detection circuit mainly comes from the low-power comparator and FPGA. The total power consumption of the packet detection module is around 34.5 uW, which is much lower than those frequency shifting system (about hundreds of uW) [24], [17].

The energy harvester on Aloba consists of a palm-size photovoltaic panel and a high-efficiency step-up DC/DC converter LTC3105 [4]. We add resistors and capacitance on board to best realize the efficiency of this converter. The power management module provides a constant 3.3V output voltage at a high power transforming efficiency (up to 1 mW output power), which is enough to afford the power consumption of the Aloba tag.

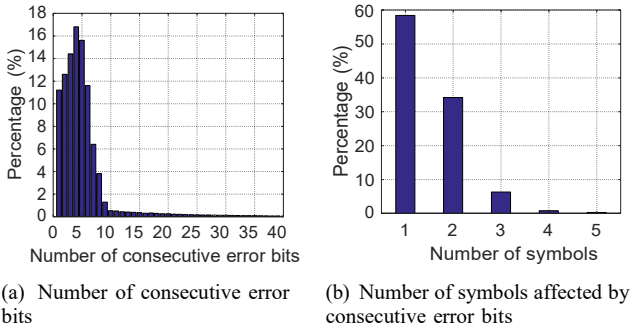

Fig. 15. The distribution of consecutive error bits in a LoRa packet.

(a) Number of consecutive error bits
(b) Number of symbols affected by consecutive error bits

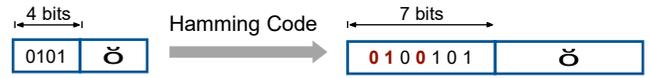

Fig. 16. Illustration of hamming coding.

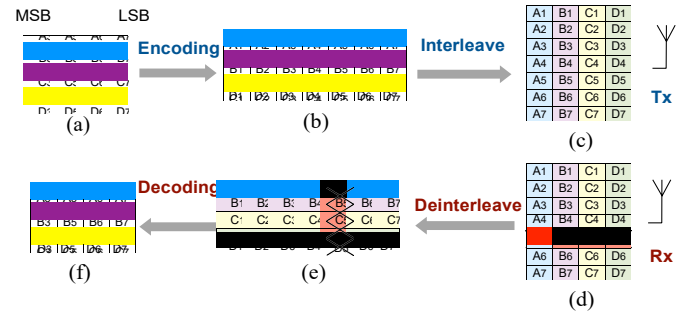

Fig. 17. The process of reliability enhancement of backscatter data decoding. (a): The backscatter data is divided into 4-bit groups. (b): Each group of 4-bit data is encoded by Hamming Code (7, 4) and is transformed to a 7-bit codeword. (c): After the interleave operation, bits in a codeword are spread across multiple symbols. (d): A whole symbol (consecutive four bits) out of seven symbols is corrupted after channel transmission. (e): After deinterleave operation, codewords are reconstructed and there is only a single corrupted bit per codeword. (f): (7, 4) Hamming decoder can correct a single error bit and obtain the backscatter data.

### B. Modulation

After detecting the LoRa preamble, Aloba waits for another 2.25 symbol times (sync. symbols) and then modulates data on the payload part of this incoming LoRa packet using OOK: reflecting the signal when transmitting a bit one, and absorbing the signal when transmitting a bit zero. The backscatter packet contains four fields as shown in Figure 14: a 16-bit baker code-based preamble $(010101\cdots010101)$, a 4-bit modulation rate field, a 8-bit payload length field, and the payload.

## V. RELIBILITY ENHANCEMENT

Due to the randomness, dynamics and burst of the channel, error bits are easy to appear during the transmission process of backscatter signals. According to the number of consecutive error bits in a LoRa packet, we classify the errors into two types: random errors and continuous errors. Random errors mean the error probability of each bit is independent, which are caused by noise or electromagnetic flash. Continuous errors mean there are multiple consecutive error bits which are caused by burst of interference. The error bits may distribute on a single symbol or multiple symbols. We conduct experiments to observe the error distribution of the decoded backscatter signals. Figure 15(a) plots the histogram of the number of consecutive error bits. We find that 95.13% of the number of consecutive error bits are less than 12. As shown in Figure 15(b), 58.4% of the error bits are distributed in a single symbol. In this section, we propose link coding mechanism, interleave operation, and coherent combining at the receivers to enhance the reliability of backscatter signal decoding.

### A. Link Coding

Hamming Code (7, 4) is used to protect the backscatter signal. As shown in Fig. 16, we add 3-bit check data to every 4-bit data and these 3-bit check data can correct 1-bit error data. The Hamming Code (7, 4) has the capability of error detection rate of $\frac{1}{4}$, which is sufficient to handle the errors of the backscatter signal.

### B. Interleave

Interleave [38] operation is robust to handle burst interference since that a codeword is spread out to multiple symbols. Corrupting a single symbol would not result in a loss of the whole codeword. The process of reliability enhancement of backscatter data is shown in Fig. 17. First, the backscatter data is divided into groups. Each group of 4-bit data is encoded by Hamming Code (7, 4) and is transformed to a 7-bit codeword. An interleaver is between the Hamming encoder and the OOK modulator, which writes in columns and reads out in rows. Because of this, bits in a codeword are spread across multiple symbols. At the receiver, the deinterleaver is the opposite, which writes in rows and reads out in columns. Although a whole symbol out of seven symbols is corrupted after channel transmission, codewords are reconstructed by these symbols and there is only a single corrupted bit per codeword. (7, 4) Hamming decoder can correct a single error bit and obtain the backscatter data.

### C. Coherent Combining

The method of coherent combining [39] at the receivers can also be used to correct the error bits of Aloba. Due to the geometric difference, the bit errors are often disjoint across different receivers. In this way, we can combine the received signals at multiple receivers to recover the Aloba packet. Moreover, the decoding complexity of this method is afforded by the LoRa receivers and it doesn't bring additional overhead to the Aloba tag.

## VI. ALOBA MAC LAYER

We sketch the MAC layer design in this section. We allocate an assigned channel among multiple Aloba tags. Each tag randomly picks up a time slot to transmit. Upon detecting the carrier signals, these Aloba tags achieve time synchronization and reflect backscatter signals. When there are multiple active LoRa nodes in the LoRaWAN, these LoRa nodes also abide by the time-domain ALOHA protocol. In this way, LoRa nodes and Aloba tags form a hybrid LoRaWAN network.

ALOHA protocol can reduce the collision probability. However, it does not guarantee the collision-free transmission. Collision happens when multiple Aloba tags select the same slot. Each tag makes its own choice independent with the other tags. If we have N tags and there are $K$ time slots, the probability that $R$ tags will be transmitted in one time slot is $P = \binom{N}{R}(\frac{1}{K})^R(1-\frac{1}{K})^{N-R}$ [40]. For instance, suppose there are 100 tags and 128 time slots, the probability that 5 tags will be transmitted in one time slot is 0.1%.

**The demodulation extension to the Multi-tag Scenario.** Our demodulation scheme can be easily extended to the multi-tag scenario. Suppose there are $M$ ($M > 1$) Aloba tags. The received signal would be the superposition of multiple backscatter signals and the carrier signal. Following the signal transformation introduced in §III-A, the receiver first transforms this received signals into a sinusoidal tone. It then follows the signal reconstruction introduced in §III-B to reconstruct the received signal. One may expect to see $2^M$ clusters on the constellation diagram. Figure 11(a) and Figure 11(b) show the constellation diagram of two tags' and three tags' replies, respectively. With those symbol clusters, we can then apply the state-of-the-art parallel decoding algorithms such as [41], [42], [43], [44] to decode the backscatter signals.

## VII. TIME SYNCHRONIZATION

There are two critical time synchronization processes that affect the system performance. We analyze the impact of synchronization accuracy in this section.

### A. The Synchronization Error on LoRa Packet Detection at the Aloba Tag

The first synchronization error happens at the Aloba tag where the tag detects the boundary of LoRa chirp and synchronizes with the payload part for modulation (backscatter). The error in synchronization will introduce a time offset $\Delta t$ between the carrier signal (start at $t_1$) and the backscatter signal (start at $t_2$), as shown in Fig. 18(a). Once the LoRa receiver detects the LoRa payload at $t_1$, the LoRa receiver set a decoding window to decode the backscatter signal. The length of decoding window is equal to the duration of one Aloba bit. During the Aloba decoding process, the signal samples with Aloba bit one and zero will aggregate into two clusters. The mismatch between the decoding window and the backscatter signal will affect the aggregation of samples within a decoding window and will further result in Aloba decoding errors.

A **moving window-based decoding strategy** can be used to tolerant the above synchronization error. As shown in Fig. 19, we select a random position as the start of the moving window and the window length is equal to the duration of one Aloba bit. As the window moves, different decoding results can be obtained. The moving position which satisfies the requirement of 16-bit baker code-based preamble of Aloba and achieves the maximum distance between two clusters with and without backscatter signals is the corresponding position of the Aloba decoding window.

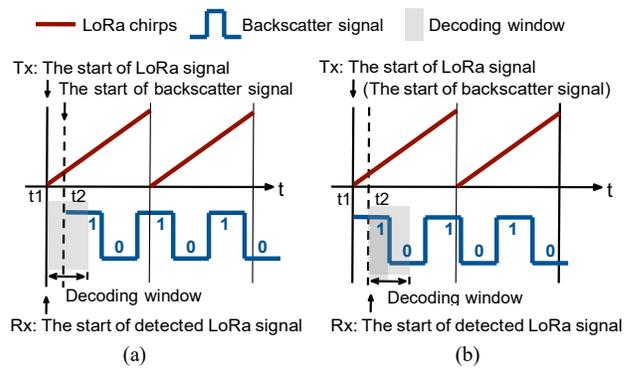

Fig. 18. Synchronization errors. (a): The synchronization error on LoRa packet detection at the Aloba tag. (b): The chirp edge detection error at the LoRa receiver.

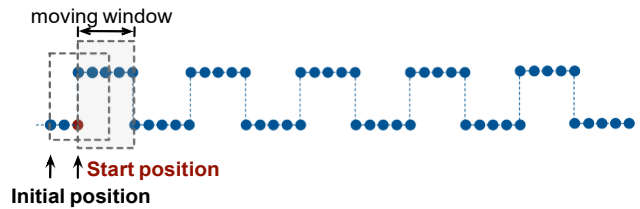

Fig. 19. The moving window-based decoding strategy. The moving position which satisfies the requirement of 16-bit baker code-based preamble of Aloba and achieves the maximum distance between two clusters with and without backscatter signals is the corresponding position of the Aloba decoding window.

### B. The Chirp Edge Detection Error at the LoRa Receiver

The second one synchronization error happens at the Aloba receiver where the receiver detects and synchronizes with the boundary of the LoRa chirp for chirp transformation. As shown in Fig. 18(b), the start of LoRa chirp at the LoRa transmitter is $t_1$, however, the detected boundary of the LoRa chirp at the LoRa receiver is $t_2$. When the time offset caused by the above chirp edge detection error is smaller than the time range bin of LoRa chirp (the minimum time offset between two LoRa chirps, which is equal to $\frac{1}{B}$), the chirp transformation can be achieved successfully and the impact on the Aloba decoding can be neglected. Otherwise, the LoRa chirps will be decoded incorrectly and the LoRa packet will be discarded at the LoRa receiver, not to mention the piggybacked backscatter data.

## VIII. IMPLEMENTATION

**Tag hardware**. The packet detection module is prototyped on a single-layer PCB using commercial off-the-shelf circuit components as shown in Figure 20(c). The packet detection module uses one omni-directional antenna with 3 dBi gain [45] and a DE0-Nano-SoC FPGA [46]. The incident signal passes through a passive SAW chip B39871B3715U410 [36] and then can be down-converted to baseband through the envelope detector. We optimize the impedance matching coefficient to provide the maximum power transfer from the antenna to the envelope detection. Finally, a low-power voltage comparator NCS2202 [37] is leveraged to quantize the output signal from the envelope detector. The packet detection module is wired to the WISP 5.0 [25] for evaluation.

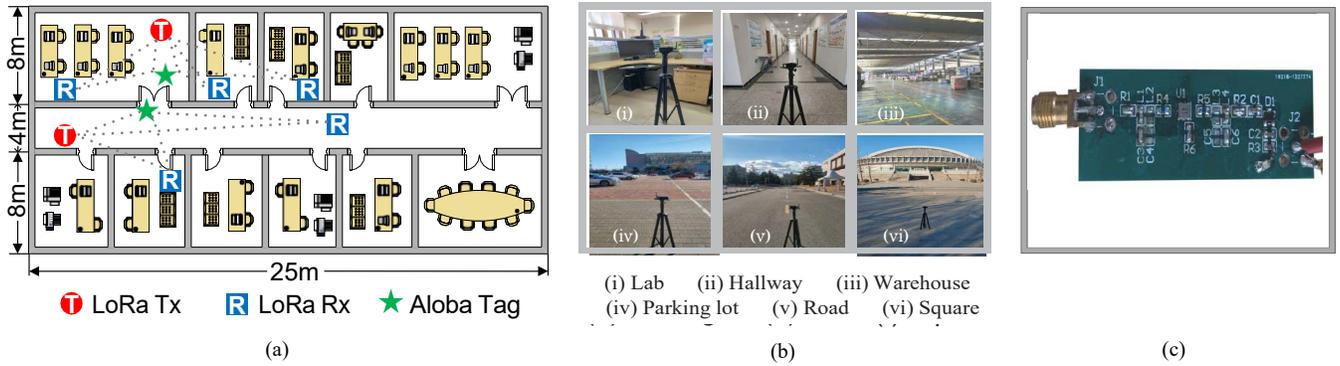

Fig. 20. The implementation of Aloba. (a): The floor plans of indoor experiment field. (b): Indoor and outdoor experiment field. (c): The packet detection module of Aloba tag.

**Transmitter and receiver**. We use a commercial LoRa node (a STM32L083RZ board [47] carrying a Semtech SX1276 [48] chip) with one 3 dBi gain omni-directional antenna [45] as the transmitter, an USRP N210 [49] equipped with the same type of antenna as the receiver. We can also choose other types of software defined radio (SDR) devices as long as they can capture the RF signals of the 900 MHz frequency band. We set the sampling rate at 10 MHz to get more sampling points and improve the robustness of the decoding algorithm based on clustering. The sampling rate only needs to meet the requirement of Nyquist Theorem, which is greater than or equal to twice the bandwidth of LoRa signal. The LoRa receiver runs the standard LoRa preamble detection algorithm [50] to detect LoRa transmissions and further locates each LoRa symbol on the payload. The receiver then runs Aloba decoding algorithm to decode both the backscatter signal and the ambient LoRa signal.

**Extension to commercial LoRa gateways**. While the current Aloba decoding algorithm is implemented on USRP, it is worth noting that this algorithm can be easily implemented on a commercial LoRa gateway, since this algorithm requires only the raw signal samples which are accessible on most LoRa RF-front, *e.g.*, Semtech SX1257 front-end [51]. We leave the algorithm implementation on commercial LoRa gateways as our future work.

## IX. EVALUATION

In this section, we first conduct head-to-head comparison with PLoRa [24], the state-of-the-art ambient LoRa backscatter system. We then conduct micro-benchmarks to study the performance of Aloba in various settings, including different LoRa bandwidth, tag-to-source distances, modulation rate, environment, and channel conditions.

### A. Experimental Setup

The LoRa transmitter and the receiver both work on channel one (902.5 MHz). The payload of each LoRa packet consists of 20 symbols. The default spreading factor (SF), coding rate, and bandwidth (BW) of the LoRa signal are 7, 1, and 125 KHz, respectively. The transmission power of the LoRa sender is 20 dBm. We evaluate Aloba both indoors (classroom, hallway and warehouse as shown in Figure 20(a).) and outdoors (open road, square and parking lot, as shown in Figure 20(b)).

We take *throughput* and *maximum backscatter range* as the key metric to evaluate Aloba's performance. *Throughput* measures the amount of backscatterred data correctly decoded within one second at the LoRa receiver. *Maximum backscatter range* refers to the maximum distance between the Aloba tag and the LoRa receiver when the bit error rate (BER) of the backscatter data is lower than 0.001. We send 1,000 LoRa packets in each experiment, and then repeat the experiment 100 times. Finally we report the averaged result to ensure the statistical validity.

### B. Head-to-head Comparison with PLoRa: Link Throughput

We compare Aloba with PLoRa [24] in various settings. PLoRa encodes one bit per LoRa symbol. The theoretical link throughput of LoRa carrier and PLoRa are $\frac{BW}{2^{SF}} \cdot SF$ and $\frac{BW}{2^{SF}}$, respectively. The theoretical link throughput of Aloba is determined by the rate of ON-OFF keying operation.

In these experiments, we place the receiver 50 m, 100 m, 150 m, 200 m, 250 m, and 300 m away from the tag. Within each distance setting, we further vary the distance between the source and the backscatter tag to measure the throughput of PLoRa and Aloba. We tune the spreading factor and bandwidth of the carrier signal to ensure the fair comparison with PLoRa. Figure 21 shows the result. We have two observations from these experimental results.

First, we observe the link throughput of Aloba is orders of magnitude higher than that of PLoRa when the LoRa receiver is within 200 m of the source (Figure 21(a)-(d)). Specifically, when the tag is collocated with the source (with an 10 cm spacing), the link throughput of Aloba is 10×4 –52.4 higher than that of PLoRa in different source-to-receiver distance settings. This is expected since the OOK design enables Aloba to tune up its throughput to best utilize the better link quality in short tag-to-source distance settings. This flexible modulation design enables Aloba to achieve even 1×–7.3 higher throughput than that of the LoRa carrier. In contrast, PLoRa adopts a fixed modulation rate and thus achieves consistently low throughput in all different distance settings.

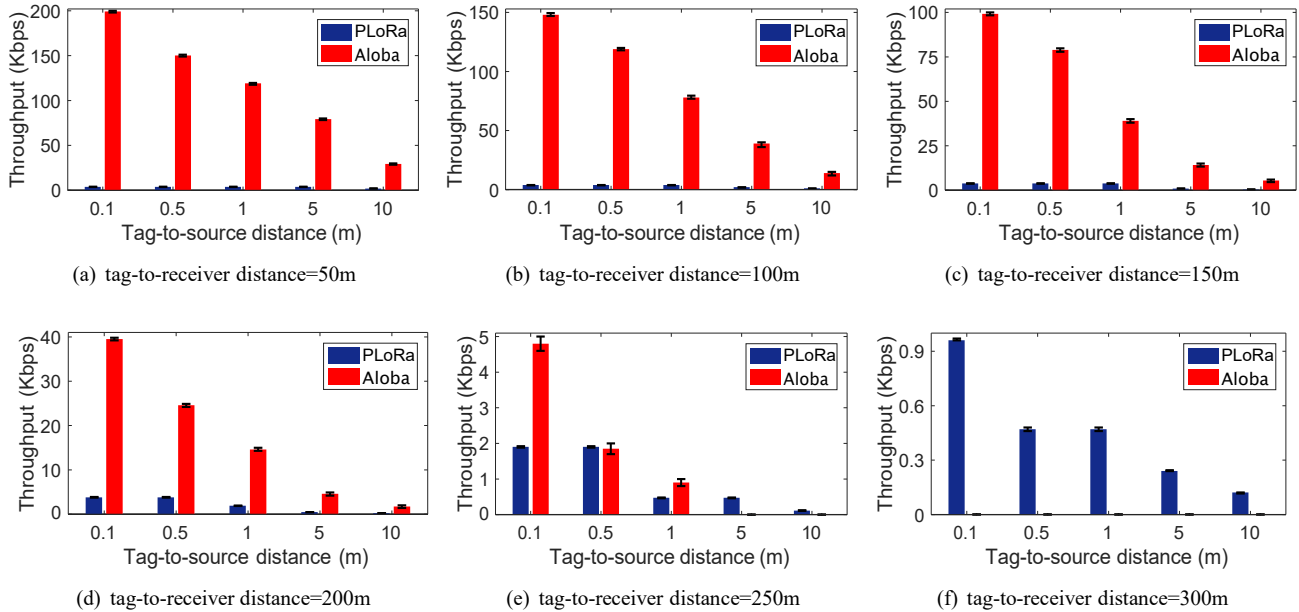

Fig. 21. Link throughput of Aloba and PLoRa in different tag-to-receiver distance settings.

Second, the link throughput achieved by Aloba and PLoRa both decreases with increasing tag-to-source distance, primarily due to the decreasing SNR of the backscattered signal (Figure 21). To expand the backscatter range, similar to the existing backscatter systems [24], Aloba has to sacrifice the throughput to ensure a longer backscatter range (Figure 21(e)). Aloba essentially relies on energy to decode backscatter signals, thus the performance gain of Aloba over PloRa is achieved mainly within short and medium communication range ($\leq$50 m shown in Figure 21(f)), since that the signal attenuation, insertion loss, and energy transformation loss on the backscatter tag result in that the backscatter signal is orders of magnitude weaker than the carrier signal. For example, Aloba achieves the throughput of 0.93 Kbps when we place the tag 1 m away from the LoRa sender and the LoRa receiver is within 250 m of the source. Other evaluation experiments about head-to-head comparison with PLoRa in different SF and BW settings can be found in SenSys version [26].

### C. Aloba Performance under Different Interference Environments

We conduct experiments to evaluate the performance of Aloba under different interference environments. In these experiments, we place the LoRa receiver 50 m away from the LoRa sender and the distance from the tag to the LoRa sender is 10 cm. An USRP N210 platform with the distance of 50 m from the LoRa receiver as the jamming generator transmits interference signals on the 902.5 MHz with the power of 20 dBm. We set the transmission interval of interference signals at 5 ms and the duration of each interference signal varies from 100 $\mu s$ to 1 ms. Given the backscatter date rate of 25 Kbps, the number of consecutive error bits varies from 4 to 40 when the duration of interference varies from 100 $\mu s$ to 1 ms. Our proposed interleave operation can theoretically disperse and correct 20 consecutive error bits. We further evaluate the performance gain brought by the link coding and interleave operation. The experimental result is shown in Fig. 22(a).

First, we observe the BER of backscatter signal increases with the increasing of the duration of interference signal. The longer the duration of the interference signal is, the more backscatter signal are affected. For example, the BER is up to 0.12 when the duration of interference signal is 400 $us$.

Second, Hamming coding (7,4) can effectively reduce the BER when there is transient jamming or interference. Hamming Coding (7,4) can correct one error bit for every seven-bit codeword. We find that the BER is reduced to 0.011 with the Hamming Coding (7,4) when the duration of interference signal is 100 $us$. Whereas, if more than 1 bit of data is wrong, the corrupted codeword can't be recovered solely by Hamming Coding.

Third, we adopt interleave operation to handle burst interference and sparse the continuous bit errors to recover more corrupted codewords. The BER is reduced to 0.001 with the interleave operation when the duration of interference signal is 400 $us$. Due to the interleave depth is limited to 20, the performance gain of interleave reduces when the duration of interference signal exceeds 500 $\mu s$ ($2 \& \frac{1}{4}$ ms). We can increase the check bit of Hamming Coding and interleave depth to further improve the anti-interference ability and enhance reliability.

Finally, we place another receiver 50 m away from the interference jammer and leverage the method of coherent combining at the receivers to correct the error bits. When the duration of interference signal varies from 600 $us$ to 1000 $us$, the BER can be significant reduced from 0.48% to 2.76%.

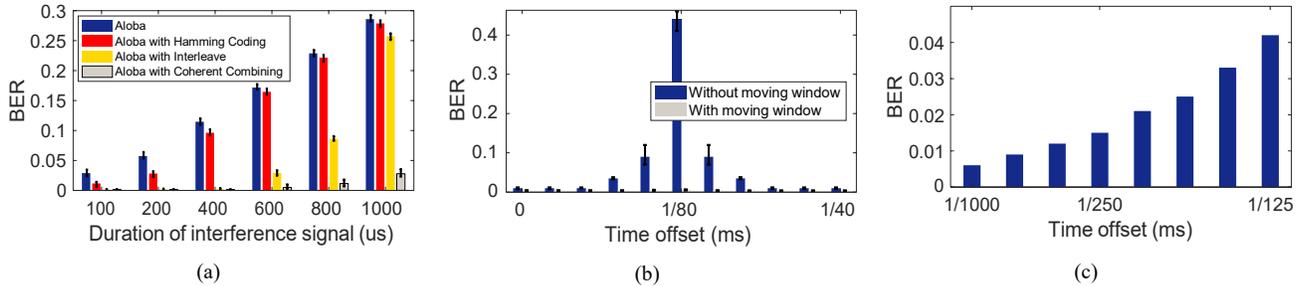

Fig. 22. Performance of Aloba in different settings. (a): Under different interference environments. (b): With time offset between the LoRa signal and the backscatter signal. (c): With the time offset of LoRa packet detection at the LoRa receiver.

*D. The Impact of Time Synchronization*

As analyzed in §VII, there are two types of synchronization errors. We conduct experiments to evaluate the impact of time synchronization on the performance of Aloba decoding. In these experiments, we place the LoRa receiver 50 m away from the LoRa sender and the distance from the tag to the LoRa sender is 10 cm. We set the switch rate of the Aloba tag is 40 KHz and the duration of one-bit Aloba backscatter data is $\frac{1}{4}$ ms. Aloba tag modulates data on the payload of LoRa signals after a waiting time, which will result in a corresponding time offset between the LoRa signal and the backscatter signal as shown in Fig. 18(a). The waiting time of Aloba tag varies from 0 to $\frac{1}{4}$ ms. The evaluation result is shown in Fig. 22(b).

First, the BER of Aloba data increases when the time offset varies from 0 to $\frac{1}{8}$ ms, half of the duration of one-bit tag data. The larger the time offset is, the more scattered the sampling points in a time window are, and the more difficult it is for these sampling points to be grouped into a cluster. Second, when the time offset is larger than $\frac{1}{8}$ ms, the decoding result will be shifted and the BER decreases. For example, the 16-bit Barker code-based preamble of one Aloba packet "010101...010101" will be shifted as "101010...101010", and the Aloba data can also be decoded. We further adopt the **moving window-based decoding strategy** to find the appropriate position of the decoding window. We observe that the BER of Aloba data is lower than 0.005 no matter what the time offset is.

Due to that it is difficult to control the time delay LoRa packet detection at the LoRa receiver shown in Fig. 18(b), we conduct an emulation experiment to evaluate the impact of chirp edge detection errors on the Aloba decoding. We suppose the time offset between the LoRa chirp at the LoRa sender and the detected LoRa chirp at the LoRa receiver varies from to 0 to $\frac{1}{12}$ ms (the time offset between two LoRa chirps). In these cases, we control the backscatter signal and the LoRa signal are aligned without time offset. The evaluation result is shown in Fig. 22(c). We find that the BER increases with the time offset of chirp edge detection. The BER is 0.042 when the time offset of chirp edge detection is $\frac{1}{12}$ ms, which indicates that the impact of chirp edge detection errors on the Aloba decoding is limited and controllable.

## X. RELATED WORK

In recent years, RF signals, such as TV, WiFi, FM, BLE, LoRa signals, have been widely exploited for backscatter communication. Ambient backscatter [11] reflects broadcast TV or cellular transmissions to achieve device-to-device communication. WiFi backscatter [12] reuses the WiFi signals to convey information by modulating the CSI and RSSI measurements. The data rate of ambient backscatter and WiFi backscatter is limited to 1 Kbps.

In order to improve the data rate, Turbo charging [13] uses the multi-antenna cancellation design with the coding mechanism to achieve the data rate of 1 Mbps. BackFi [15] modulates information by changing the phase of the received WiFi signals, which improves the communication rate to 5 Mbps. Passive WiFi [12] enables a passive tag to generate 802.11b transmissions by leveraging a dedicated excitation device. HitchHike [17] allows a backscatter tag to embed its information on standard 802.11b packets, by translating the original transmitted 802.11b codeword to another valid codeword. FreeRider [18] extends the technique of codeword translation to other radios, such as 802.11g/n, Bluetooth, and ZigBee. OFDMA-WiFi [21] enables OFDMA in WiFi backscatter for capacity and concurrency enhancement. But the farthest communication range is only tens of meters.

To further improve the communication range, researchers focus on Low-Power Wide-Area Network (LPWAN) technologies. Among the LPWAN technologies [52], [53], LoRa [50] is resilient to interference due to its high receiving sensitivity, making it a natural choice for backscatter. LoRa backscatter [23] synthesizes legitimate LoRa packets to extend the communication to 2 km. However, it requires a dedicated device to generate the excitation signal. PLoRa [24] is the most relevant work with Aloba, which takes ambient LoRa transmissions as the excitation signals and modulates the original LoRa chirp signal into a new standard LoRa chirp signal at another frequency band. Whereas, PLoRa is not spectrum efficient and inevitably consumes the already crowded wireless spectrum.

Compared to the existing works, Aloba adopts the modulation of ON-OFF Keying (OOK) and the data rate can be easily adjusted by tuning the frequency of this RF switch. By taking the ambient LoRa signals as the excitation, the backscatter tag could leverage the unique processing gain brought by the chirp signal design to enable long-range backscatter communication. In this way, Aloba supports flexible data rate at different transmission range. Moreover, Aloba achieves high spectrum efficiency.

## XI. DISCUSSION

### A. The Impact of LoRa Duty Cycle on Aloba

We propose Aloba to enhance rather than to replace the LoRa communication. Indeed, the Aloba tag relies on the LoRa signals as the carrier signals to transmit data. Multiple Aloba tags join a LoRaWAN to form a hybrid LoRaWAN network. Aloba provides battery-free but efficient communication, which is especially suitable for periodical sensing applications, where the sensor nodes typically work in low-duty cycle mode but desire relatively high data-rate communication. Moreover, with increasingly deployed LoRa nodes in the environment [54], one may expect to see increasing space to deploy Aloba tags and utilize LoRa signals therein for backscatter.

### B. The Adjustment between Data Rate and Communication Range

There are three potential solutions to achieve the tradeoff between data rate and communication range. First, the basic idea is to flexible set the data rate offline according to the geographic location during the deployment of the Aloba tag. Second, the packet detection module of the Aloba tag provides the RSSI information of the LoRa carrier signal, which can be used as an indicator to infer the communication range between the Aloba tag and the LoRa receiver, according to the signal attenuation model [22]. According to the inferred communication range, the Aloba tag can set the corresponding data rate. Third, we may replace the existing packet detection module with Saiyan [55], which allows the Aloba tag to demodulate the incident command/feedback LoRa signals. As a low-power component, Saiyan can serve as a plug-in module to directly benefit Aloba without much engineering efforts. On the software side, we only add a decision layer to configure the data rate based on the decoded messages.

### C. The SNR Requirement of Aloba Decoding

Aloba takes the ambient LoRa transmissions as the excitation and piggybacks the in-band OOK modulated signals over the LoRa transmissions. There is a SNR gap in between at which LoRa symbol is decoded but the Aloba symbol cannot be decoded. Specifically, the chirp-modulated LoRa signals can be decoded at -30 dB SNR, while the OOK-modulated Aloba signals can be decoded at 0 dB SNR. The SNR difference results in that the Aloba tag cannot achieve similar communication range as the active LoRa nodes. There are multiple ways to increase the backscatter range. For example, leveraging beamforming techniques or negative impedance components like tunnel diode and we leave it as our future work.

### D. The Impact of Vibration, Rotation, and EMI

There are interference of vibration, rotation, EMI introduced by the industrial environment. First, the influence of periodic vibration and rotation of industrial machine on carrier signal is limited and can be ignored. On the one hand, periodic vibration and rotation produce mechanical wave, which is different from electromagnetic wave of carrier signal. Hence, the periodic vibration and rotation don't introduce a period change on the carrier signal. On the other hand, when the Aloba tag attached to an industrial machine, the vibration and rotation of the machine may slightly change the path difference between the backscatter signal of Aloba and the carrier signal of LoRa. Whereas, the influence of the fluctuation of path difference is so small that it is generally negligible. For example, the typical values of vibration period and amplitude of an industrial machine are 1 KHz and 200 $\mu$m, respectively [8]. That is to say, the fluctuation of path difference is approximately 200 $\mu$m. According to the equation of $\varphi = \frac{d}{\lambda} 2\pi$ (where $\varphi$ is phase, $d$ is path difference, $\lambda$ is wave length), the phase change caused by the fluctuation of path difference fluctuation is only $3.76 \times 10^{-5}$, which can be ignored. Second, other electromagnetic interference (EMI) may cause the flicker or disturbance on the carrier signals. We enhance the anti-interference ability of Aloba by leveraging channel coding mechanism, such as Hamming Coding and Interleave Operation.

## XII. CONCLUSION

Aloba is an ambient LoRa backscatter design using ON-OFF Keying that provides flexible data rate and transmission range for different IoT applications and deployments. By allowing the coexistence of the backscatter signal and the carrier signal in the same frequency band, Aloba achieves a higher spectrum efficiency. Our design contributions are a low-power backscatter design that can pick up the ambient LoRa transmissions from other interfering signals and a decoding algorithm running on the LoRa receiver that can decode both the backscatter signal and the LoRa excitation signal from their superposition. We propose link coding mechanism and interleave operation to enhance the decoding reliability. We also discuss the impact of synchronization on the Aloba performance and leverage moving window-based decoding strategy to tolerant synchronization errors. Evaluation results demonstrate that Aloba can achieve various data rates (39.5–199.4 Kbps) at various distances (50–200 m) in the wild. Compared with the state-of-the-art system PLoRa [24], Aloba is 10.4–52.4× better in terms of throughput.

## ACKNOWLEDGMENT

This work is supported in part by National Key R&D Program of China No. 2017YFB1003000, National Science Fund of China under grant No. 61772306, the Smart Xingfu Lindai Project, and the R&D Project of Key Core Technology and Generic Technology in Shanxi Province (2020XXX007).

JOURNAL OF LATEX CLASS FILES, VOL. 14, NO. 8, AUGUST 2021 13

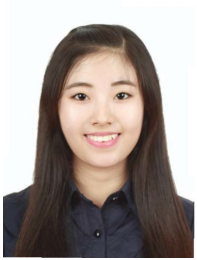

**Xiuzhen Guo** received the B.E. degree in the School of Electronic and Information Engineering from Southwest University in 2016. She is currently a PhD student in Tsinghua University. Her research interests include Internet of Things and wireless networks.

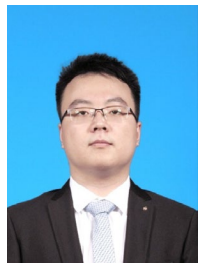

**Haotian Jiang** is currently an undergraduate student in Tsinghua University. His research interests include wireless network co-existence and cross-technology communication.

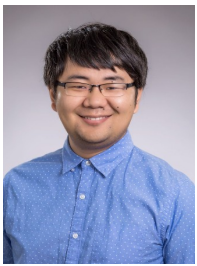

**Longfei Shangguan** is a senior researcher at Microsoft Cloud & AI, Redmond. He received his B.E. degree in Xidian University, and his PhD degree in Hong Kong University of Science and Technology. His research interests include networking, IoT, and wireless systems.

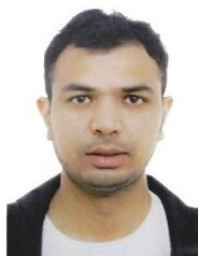

**Awais Ahmad Siddiqi** earned his BS degree in Electronics Engineering from University of Wah, and MS degree from Northeastern University, China. Currently, he is enrolled as a PhD student in Tsinghua University. His research interests include Internet of Things, remote sensing and wireless networks.

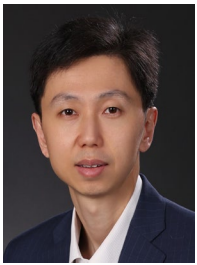

**Yuan He** is an associate professor in the School of Software and BNRist of Tsinghua University. He received his B.E. degree in the University of Science and Technology of China, his M.E. degree in the Institute of Software, Chinese Academy of Sciences, and his PhD degree in Hong Kong University of Science and Technology. His research interests include wireless networks, Internet of Things, pervasive and mobile computing. He is a senior member of IEEE and a member of ACM.

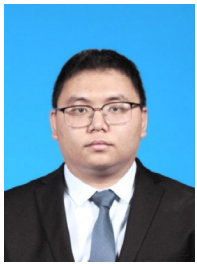

**Jia Zhang** is currently an undergraduate student in Tsinghua University. His research interests include wireless sensor networks and cross-technology communication.

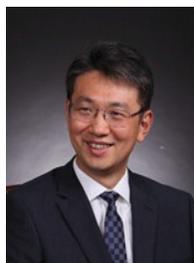

**Yunhao Liu** received his BS degree in Automation Department from Tsinghua University. He received an MS and a Ph.D. degree in Computer Science and Engineering at Michigan State University, USA. Yunhao is now MSU Foundation Professor and Chairperson of Department of Computer Science and Engineering, Michigan State University, and holds Chang Jiang Chair Professorship at Tsinghua University. He is an ACM Distinguished Speaker and now serves as the Editor-in-Chief of ACM Transactions on Sensor Networks. His research interests include sensor network and pervasive computing, peer-to-peer computing, IOT and supply chain. Yunhao is a Fellow of IEEE and ACM.